\documentstyle[aps,prl,twocolumn,epsf]{revtex}
\begin{document}
\title{Statistics of DNA sequences: a low frequency analysis}
\author{Maria de Sousa Vieira\cite{email}}
\address{Department of Biochemistry and Biophysics, University of California,
San Francisco, CA 94143-0448.}
\maketitle
\begin{abstract}
We study statistical properties of DNA chains of thirteen 
microbial complete genomes. We find that the power spectrum of 
several of the sequences studied flattens off in the low 
frequency limit. This implies the correlation length in 
those sequences is much smaller than the entire DNA chain.  
Consequently, 
in contradiction with previous studies, 
we show that the fractal behavior of DNA chains not always  
prevail through the entire DNA molecule. 
\end{abstract}
\narrowtext
\section{Introduction}\label{s1}
The statistics of DNA sequences is an active topic of research 
nowadays. 
There are studies on the power spectral density, random
walker representation, correlation function\cite{li2}, etc.
Although some of the studies are in contradiction with each other, there
is a consensus with respect to the reported behavior of the power spectrum
of DNA sequences. For high frequencies it is roughly flat, with a
sharp peak at $f=1/3$, which has been shown to be due to nonuniform
codon usage\cite{grosse,lee}. For smaller frequencies, it has been reported that
it presents a power-law
behavior with exponent approximately equal to $-1$, that is, $1/f$ noise. 
Since a cutoff of the power-law exists at high frequencies,
it has been called ``partial power-law"\cite{li}.
The presence of 
``1/f" noise in a given frequency interval indicates the 
presence of a self-similar (fractal) structure 
in the corresponding range of wavelengths, whereas a 
flat power spectrum indicates absence of correlations (white noise). 

It is an important question to know whether or not the power-law
behavior of the power spectrum of a given DNA chain extents up to 
the smallest frequencies.
If this occurs, it would imply
that the fractal behavior of that DNA chain spans to  
the entire chain, 
and  
that the correlation length of 
the chain is not smaller than the chain size.
Some studies have claimed that the fractal behavior of
DNA prevails through the entire DNA molecule\cite{lu}. The
aim of this paper is to show that 
this is not generally correct. 
                                               
We have done  statistical analysis of  the DNA of thirteen microbial 
complete genomes\cite{genome},  
that is,   
{\it Archaeoglobus fulgidus} (2178400 bp), {\it Aquifex aeolicus} (1551335 bp), 
{\it Bacillus subtilis} (4214814 bp), {\it Chlamydia trachomatis} (1042519 bp), 
{\it Escherichia coli},  also known as Ecoli (4639221 bp), 
{\it Treponema pallidum} (1138011 bp), {\it Haemophilus influenzae Rd}
(1830138 bp), 
{\it Helicobacter pylori 26695} (1667867 bp),  {\it Mycoplasma pneumoniae} 
(816394 bp), 
{\it Mycobacterium tuberculosis H37Rv} (4411529 bp), 
{\it Pyro-h Pyrococcus horikoshii OT3} (1738505 bp),
{\it  Synechocystis PCC6803} (3573470 bp), and
{\it Mycoplasma genitalium G37} (580073 bp).
We have found that the  behavior of power spectrum at
small frequencies can be different for different organisms. Also, it
can be different for different nucleotides in the same
organism.
Thus, for some organisms, the behavior of the power spectrum (PS) as 
a function of the frequency shows, in a log-log plot, three different 
regions, instead of two, reported previously\cite{li,lu,voss}.   
That is, as the frequency increases, it changes from (on average) 
a flat function, a power-law, and then flat again\cite{note}, showing that  
the fractal structure of DNA sequences not necessarily extends up 
to the total length of the chain.  
The flattening of the power
spectrum at low frequencies is just a signature of the fact that the
correlation length of DNA sequences is, for many sequences, much smaller
than the entire length of the DNA chain. 
We have calculated the autocorrelation function (AF) of the 
nucleotides in the DNA chains of the organisms  mentioned above. 
We have found that 
in some of the organisms   
the correlation length is  of the order of a few thousand
base-pairs. In others, the correlation length is very large, being not
smaller than 100,000 base-pairs. 

A DNA chain is represented by a sequence of four letters, 
corresponding to four different nucleotides: adenine (A),
cytosine (C), guanine (G) and thymine (T).  
The calculation of the power spectrum or the autocorrelation function
requires that this symbolic sequence be transformed into a numerical
one. Several methods have been proposed for this\cite{li,voss,stanley}. 
Here we use the method introduced by Voss\cite{voss}, which has been
shown in \cite{coward} to be equivalent to the method used in\cite{li}.
In Voss's method one 
associates 0 to the site in
which a given symbol is absent and 1 to the location where it is
present. So, for a given DNA sequence there will be four different
numerical sequences, corresponding to the sequences 
associated with A, C, G and T. In his original paper, Voss calculated the PS
for each one of these sequences and summed them 
to find the average PS.
Here, we treat them  distinctly,   
because we also want to know about the similarities and differences
of the statistical features of different nucleotides in a given  
DNA sequence.
 
By artificially linking flank sequences together,  
Borstnik {\sl et al.}\cite{borstnik} 
found a behavior for the PS as a function of the frequency that 
was flat, then an exponential decay, then flat again. Our studies 
of {\sl complete} sequences show that the behavior of the PS does 
not show any exponential decay in the region of 
intermediate frequencies. We found instead a 
power law. However, for low frequencies we also find 
a flat PS in several of the sequences studied. A flat PS  
at high frequencies is observed in all cases. 

\section{Statistical Analysis}\label{s2}

\subsection{Power Spectrum}\label{ss1}

Let us use Voss's method\cite{voss} and 
denote by $x^A_j$ the numerical value associated with the symbol A.  
Then one has 
$x^A_j=1$ if symbol $A$ is present at location $j$ and $x^A_j=0$ otherwise. 
Similar transformation is made for symbols C, G and T. 
Consequently, 
the DNA can be divided into four different binary subsequences of 0's and 1's, 
associated with the symbols A, C, G, and T.  

The Fourier transform of a numerical sequence ${x_k}$ of length $N$ 
is by
definition,
\begin{equation}
V(f_j)\equiv {{1}\over{N}} \sum _{k=0}^{N-1} x_k \exp(-2\pi ikf_j),
 \label{eq1}
\end{equation}
where the frequency $f_j$ 
is given by $f\equiv j/N$, and $j=0,...,N-1$. 
The PS is defined as
$S(f_j)=V(f_j)V(f_j)^*=|V(f_j)|^2$. From the definition, we can see that 
$V(f_0)=<x_k>$, where the brackets denote average along the chain. Consequently, 
this quantity carries no information about the relative positions of the 
nucleotides. Because of this, we usually neglect this quantity in our 
calculations, that is, we concentrate only on frequencies   
with $j>0$.  

Since DNA sequences have a large number of base-pairs, 
and the PS presents considerable fluctuation,  
some kind of averaging is 
usually done  to plot this quantity as a function of the frequency. 
The way of averaging done so far is
the following 
\cite{li,lu,voss}:
the DNA chain of length $N$ is divided into non-overlapping
subsequences of length $L$. Then, the power spectrum of each of these
segments is computed and averaged over the $N/L$ subsequences. 
In this method the smallest frequency for which the PS can be 
calculated is,  
of course, $f=1/L$. Consequently, the behavior of frequencies 
in the range $[1/N,1/L]$ 
is unknown. 
An
example of such a calculation for  Ecoli
is shown in Fig.~\ref{f1}, where the DNA chain was divided in
subsequences of 8192 nucleotides. 
A clear power law, followed by an approximate flat region 
with a sharp peak at $j=1/3$, is seen. 
To  avoid overlap of the curves, we have displaced the PS of 
cytosine, guanine and thymine  by dividing it  
by  $10$, $10^2$ 
and $10^3$, respectively. 
Since the power spectrum for sequences of real numbers 
is symmetric with respect to the axis $f=0.5$, we plot only 
the PS for frequencies in the interval of  $0$ to $0.5$. 
A similar  figure for adenine  is shown in 
\cite{lu}.


In this paper we show that another way of averaging allows  
one to calculate the PS for smaller frequencies than the method 
described above, and then verify what happens to the power-law 
as the frequency decreases. 
More specifically, we calculate the 
mean PS in a sliding window of $n$ points, 
with adjacent windows having an overlap of $n-1$ points.   
The average PS in  
each window will determine the values of the smoothed resulting sequence.    
In mathematical terms we can express this as     
\begin{equation}
\overline S(f_j)={{1}\over{n}} \sum _{m=j-\Delta} ^{j+\Delta} S(f_m)
\label{eq5}
\end{equation}
where $\Delta=(n-1)/2$,   
$n$ is taken an odd number, and $j$ varies from $\Delta +1$ to $N-\Delta-1$. 
Although the new sequence in this 
method is smoother than the original one, its length 
is only smaller than it by 
$2\Delta$ points.
We have found that this method shows the same behavior for 
moderate and high frequencies as 
the method used in\cite{li,lu,voss}. However, it much superior for studies at 
low frequencies. 

To speed up the calculations of the PS we have used, as it is normally done, 
the Fast Fourier Transform algorithm\cite{nr}. This  
algorithm speeds up the calculation of the PS by 
a factor of $N/log_\alpha N$, but it  requires that length 
of the sequence analyzed be an  
integer power of the integer $\alpha $, which usually is taken 
to be two. Since the length of DNA sequences are not generally 
equal to an integer power of two, we take in our computation the  
largest subsequence, starting from the beginning of the chain, 
that fulfills this requirement. More 
specifically, we take the first $N'=2^K$ nucleotides, where 
$K$ is the largest power of 2 satisfying the requirement that 
$N' \le N$, with $N$ being the total size of the DNA chain. In 
this way, the number of nucleotides not included in the calculation 
is always smaller, and in many cases much smaller, than $N/2$. 
We have also done calculations considering the entire length of 
the DNA and zero padding the sequence to the next integer power of 2, 
as described in\cite{nr}. The results remain essentially the 
same as the ones we show here. 

Since our method shows the same behavior for the PS in 
the range of intermediate and large frequencies as the other 
averaging method, and 
also due to the large size of the DNA chains,  
we plot the PS only in the frequency range $[1/N,0.01]$.
We show in Fig.~\ref{f2} the results of our calculation for 
$n=33$ for four representative cases of the thirteen ones studied. 
For clarity, we have displaced the PS of 
C, G and T by dividing it 
by  $10$, $10^2$ 
and $10^3$, respectively. 
In this way, an overlap of the curves is avoided. 
Our results show that the low frequency  PS associated with each 
of the nucleotides in the organisms studied fall into one of 
the following cases:

(a) All the four PS associated with the four different 
nucleotides 
flattens off at low frequencies. In these cases there are
three regions in the PS versus frequency curve.  
At both low and high frequencies the PS is 
of white noise type and the middle regions is characterized 
approximately by a power-law behavior, that is, in a log-log plot the 
PS satisfy $S \sim f^{-\gamma}$, with $\gamma > 0$. 
This is for example the 
case of Ecoli, shown in Fig.~\ref{f2}(a). When compared with 
Fig.~\ref{f1} or with Fig. 1 of \cite{lu}  we see that 
the averaging method of\cite{li,lu,voss} does
not show the true behavior of the PS at low frequencies.  
In this calculation  we used the first $2^{22}$ 
nucleotides, which corresponds to $90\%$ of the Ecoli DNA.
We show another case with the same behavior in Fig.~\ref{f2}(b), which 
is the PS of {\sl Aquifex aeolicus}.
For the PS of {\sl  Aquifex aeolicus} we used 
the first $10^{20}$ sites, which corresponds to 
$68\%$  of the chain length.
The other organisms, 
among the ones studied, that show the same PS behavior are
{\sl Archaeoglobus fulgidus},  {\sl Synechocystis PCC6803},   
{\it Mycoplasma pneumoniae},  and 
{\sl Mycobacterium tuberculosis}. 

(b) The second type of behavior is the one in which
the PS at small frequencies of all the nucleotides presents 
a power law behavior, which is approximately an extension of the 
PS behavior at intermediate frequencies. For these 
organisms, the PS presents only two regions: a flat 
one at high frequencies, and a power law behavior for 
intermediate and low frequencies. 
A typical case for this kind of behavior 
is shown in Fig.~\ref{f2}(c), which is the PS of 
{\it Bacillus subtilis}. 
In the calculation of the PS in this case we have used the
first $10^{22}$ sites, which corresponds to 
$99\%$ of the total length of the chain.
The other organisms studied 
that have similar PS are: {\it Treponema pallidum}, 
{\it Pyro-h Pyrococcus horikoshii OT3}, 
and {\it Mycoplasma genitalium}. 

(c) The third, and last, type of behavior we have seen 
is the one in which, for a given organism, different 
nucleotides present different asymptotic behavior for 
the PS at low frequencies. That is, the PS  
flattens off for some of the nucleotide sequences, 
and for the others it remains approximately a power-law. 
An example of such a behavior is shown in 
Fig.~\ref{f2}(d), which is the PS of 
{\it Haemophilus influenzae Rd}. 
We see that different behavior for the PS are  
grouped in pairs. In all the cases studied we found that  
the PS of A is qualitatively similar to the PS of T and the one of 
C is similar to the one of G.
This kind pairing of the statistical features of 
nucleotides has been reported for yeast chromosomes in
\cite{li2}. This is probably caused by the 
strand symmetry of DNA sequences, reported in\cite{strand}.
In the calculation of the PS we have used 
the first $10^{20}$ sites of the DNA chain, which 
corresponds to $57\%$ of the total number of nucleotides.
Since a large number of sites are left out of the 
calculation, we have also analyzed the PS of the 
central and final region of the chain. We verified 
that the results  remain essentially the same as the 
ones shown in Fig.~\ref{f2}(d).
The other organisms that have similar statistical 
features for the PS are {\it Chlamydia 
trachomatis} and {\it Helicobacter pylori 26695}.


\subsection{Autocorrelation Function}\label{ss2}

The autocorrelation function  R(l) of a numerical sequence is, by definition,
\begin{equation}
R(l)=< x_k x_{k+l} >,
\label{eq2}
\end{equation}
where the brackets denote  average over the sites along the chain. 
For $l=0$, Eq.~(\ref{eq2}) implies $R(0)=<x_k^2>$, which is a
quantity carrying no information about the  relative position of 
the nucleotides. As in the case of the power spectrum for $S(0)$, this 
quantity will be neglected in our calculations.

Statistical independence between sites separated a by distance $l$ implies 
that 
$< x_k x_{k+l} >=< x_k>^2$.
The value of $l$ above which this condition is satisfied (on average) is 
called the correlation length. 
DNA molecules, depending on the organism, can form an 
open or a closed loop. 
Bacterial DNA usually forms a closed loop\cite{circular}. 
For circular chains, the autocorrelation function  and the PS form Fourier 
transform pairs (this is the Wiener-Khintchine theorem)\cite{comment1} .   
In order to consider the entire DNA sequence (without having 
the constrains of the Fast Fourier algorithm) we calculate the 
AF using its plain definition, that is, Eq.(\ref{eq2}), and 
not via Fourier transforming of the PS\cite{comment2}. We present 
results for $l$ in the interval  $[1, 10^5]$.
This is a much  larger interval than the ones considered 
in previous publications\cite{li2,grosse2}, which took $l$ in  $[1, 10^3]$. 
It is obvious that 
when $l << N$, as it occurs here,  
it does not matter if we consider open 
or closed boundary conditions. Since we find computationally 
easier to consider open boundary conditions, we present the 
results of the AF for this case.  
It is beyond the scope of this paper to study 
cross-correlation between two different kinds of nucleotides.  
Such a kind of study can be found for example in \cite{li2,grosse}.

We show in Fig.~\ref{f3}  the AF versus $l$ for the sequences whose PS we 
displayed in  Fig.~\ref{f2}. Since the AF presents a strong  oscillation 
of period 3\cite{grosse}, 
we chose $n$ to be a multiple of 3 in order to smooth it out. 
Here we have used $n=33$ (there was no particular reason 
for choosing $n$  a multiple of 3 in the calculation of the PS). 
In  Fig.~\ref{f3} the horizontal lines 
are the corresponding values of  $< x_k>^2$. When 
$R(l)\equiv <x_k x_{k+l}> \approx < x_k>^2$ statistical independence 
between the nucleotides of a given type holds.  
As  Fig.~\ref{f3} shows, when $l \lesssim 100$ the AF is roughly flat    
for some sequences, and for others it is approximately a power-law
\cite{powerlaw}.
Then, as $l$ increases we see a regime of a power-law in all cases.  
For the interval of $l$ studied, we observe that the AF can get  
flat again as $l$ increases even more (with $R(l)\approx < x_k>^2$), or   
not reach a plateau. 
For the sequences where the PS flattens off at low frequencies, we 
expect that the AF will get flat for larger $l$, with statistical 
independence holding. However, for most of the cases studied, this 
happens when $l >> 10^5$. Only the  
AF    
of  {\sl Aquifex aeolicus} 
seems  to reach a plateau for $l$ in the interval $[1, 10^5]$ for 
all the nucleotides.  
This is shown in Fig.~\ref{f3}(c) and Fig.~\ref{f3}(d), where 
we observe that the correlation lengths for this organism appear to be between 
$10^3$ and $10^4$. 
For the 
other organisms studied, we see a wide variety of behaviors  
for the AF in the region of $l\in [10^3, 10^5]$. 
As  Fig.~\ref{f3}  shows, we find cases in which the 
AF reaches a plateau with statistical independence between 
the nucleotides, in others we see a slow decrease of the AF, such 
as the AF of A for {\sl Bacillus subtilis}. We also find 
an abrupt change of slope in a plateau region, 
like the AF for A and of G  in 
{\sl Haemophilus influenzae Rd}. And most interestingly, we 
find the presence of anti-correlations, that is, 
$<x_k x_{k+l}>$ being smaller than  $<x_k>^2$. 
This implies that sites separated by 
a given distance tend to be occupied by different nucleotides.  
The case in which this appears 
more strongly is in the AF of C for {\sl Haemophilus influenzae Rd}.     
We have also observed that most of the sequences present a peak 
in the AF at $l\approx 100$. The reason for this is unknown to 
us. 

\section{Conclusion}\label{s3}
In summary, 
we have studied statistical properties of the complete DNA 
of thirteen microbial genomes and shown that its fractal   
behavior not always prevails through the entire 
chain. For some sequences the power spectrum gets flat at low frequencies, and 
for others it remains a power-law.  
In the study of the autocorrelation function we have found a rich 
variety of behaviors, including the presence of anti-correlations.   

\acknowledgments
This paper is an outgrowth of work done with 
H. J. Herrmann, to whom I am grateful for introducing me in this 
subject. 

\begin{figure}
\caption[f1]{
Power spectrum of Ecoli calculated by dividing the entire 
DNA chain in subsequences of 8192 nucleotides. The curves for 
C, G and T have been multiplied by factors of $10^{-1}$,  $10^{-2}$ and 
 $10^{-3}$, respectively, to avoid overlap of the curves.  
}
\label{f1}
\end{figure}

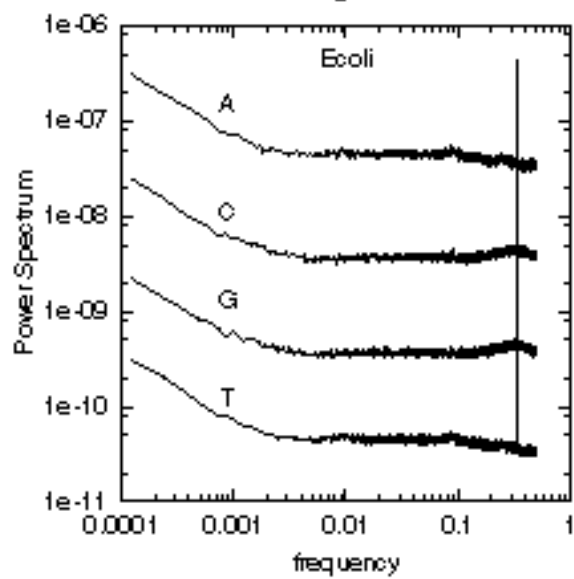
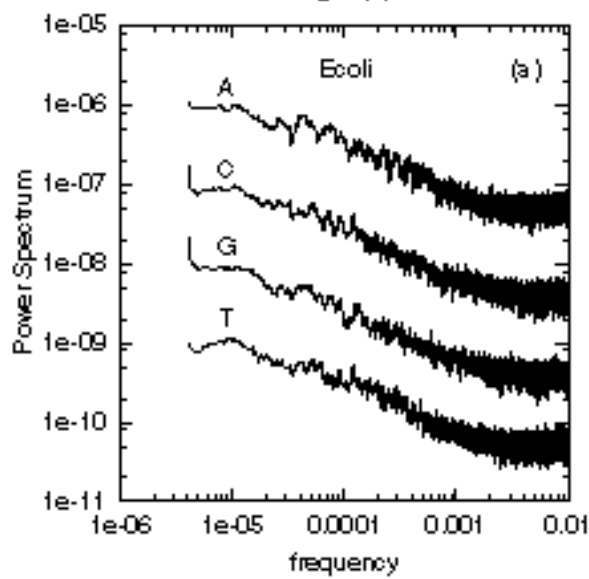
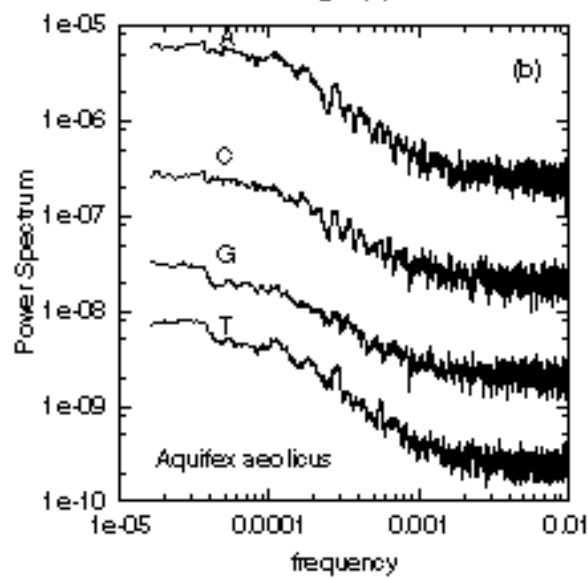
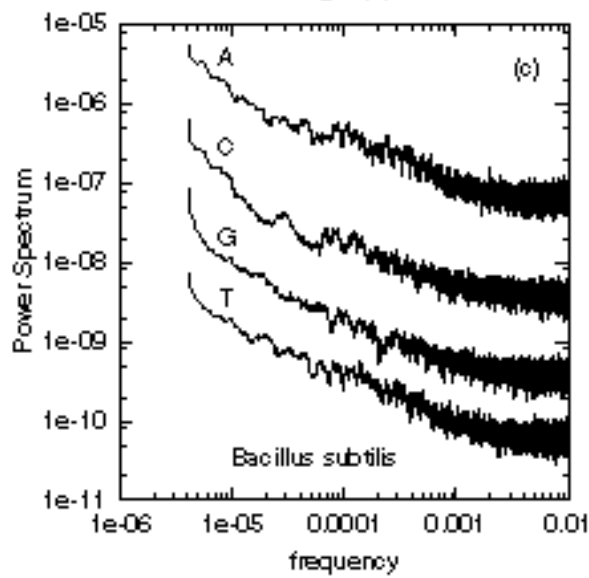
\begin{figure}
\caption[f2]{
Power spectrum  of (a) Ecoli, (b)  
{\it Aquifex Aeolicus}, (c) {\it Bacilus subtilis} and (d) 
{\it Haemophilus influenzae Rd}.  
The curves for   
C, G and T have been multiplied by factors of $10^{-1}$,  $10^{-2}$ and 
 $10^{-3}$, respectively, to avoid overlap of the curves. 
}
\label{f2}
\end{figure}

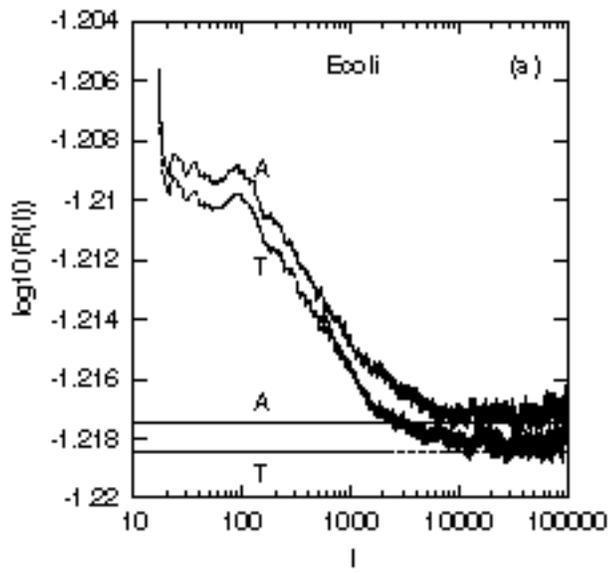
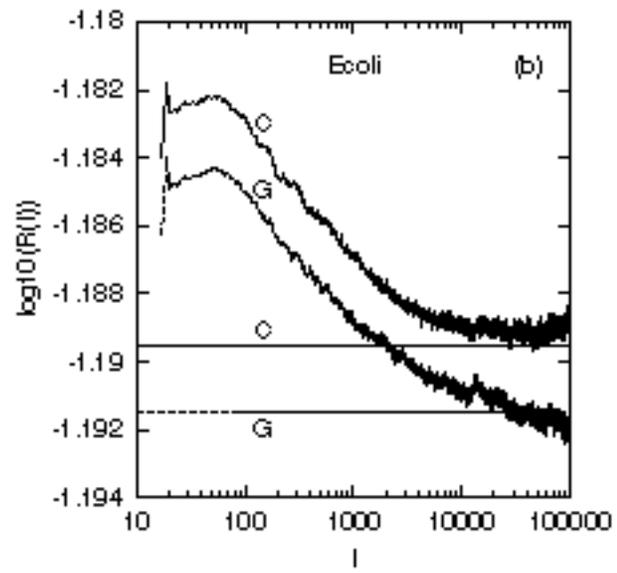
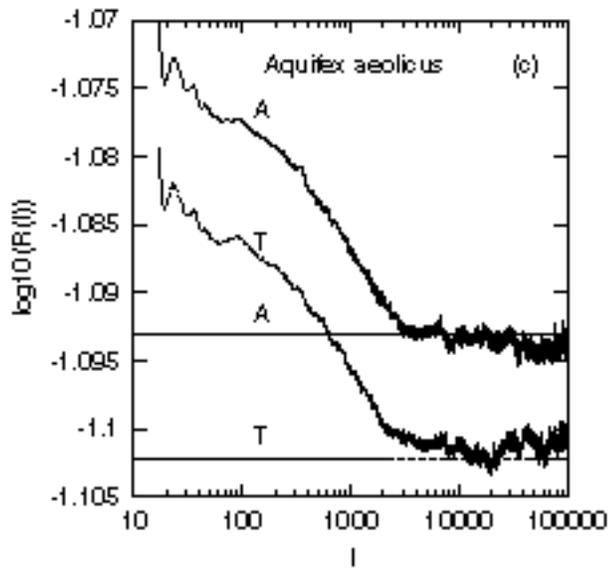
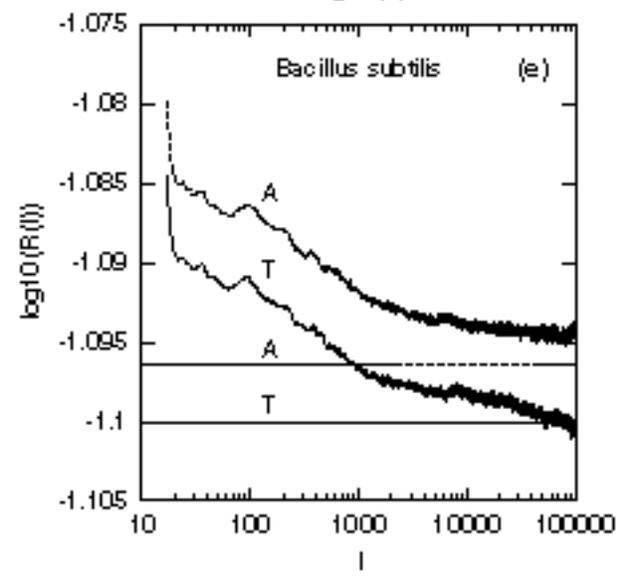
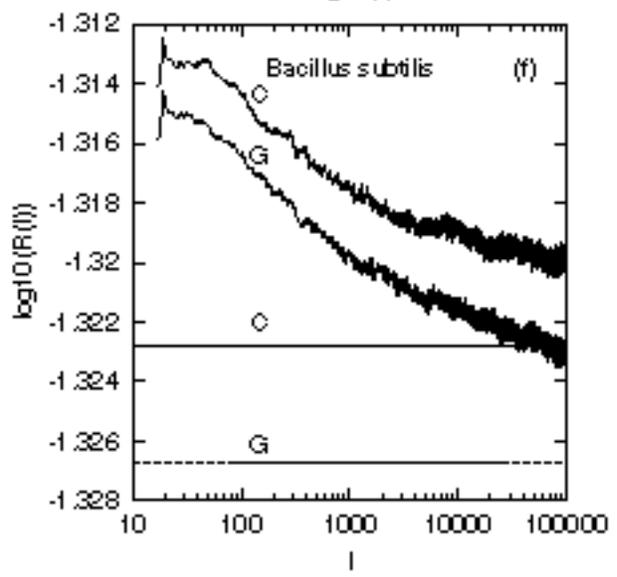
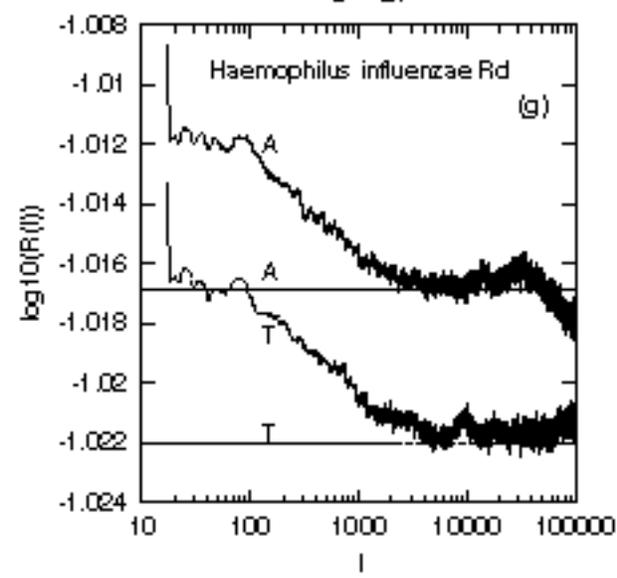
\begin{figure}
\caption[f3]{
Autocorrelation function of Ecoli for (a) A and T, (b) C and G, 
{\it  Aquifex Aeolicus} for (c) A and T, (d) C and G,   
{\it Bacilus subtilis}  for (e) A and T, (f) C and G, and 
{\it  Haemophilus influenzae Rd}  for (g) A and T, (h) C and G.  
The horizontal lines are $<x_k>^2$. Statistical linear 
independence between the nucleotides in  a given sequence occurs when 
$R \equiv <x_kx_{k+l}>=<x_k>^2$. 
}
\label{f3}
\end{figure}



\begin{references}
\bibitem{note} Note that when we 
say ``a flat region" for high frequencies we are also implying 
the existence of well known peaks in that region, 
the highest one being the 
peak at $f=1/3$.
\bibitem{comment1} One can easily show that for  
free boundaries the PS and AF are related by  
$S(f_j)=R(0)+2\sum_{k=1}^{N-1} (1-k/N)R(k)\cos(2\pi k f_j)$, 
showing that these two quantities are not Fourier transform 
pairs. However, if the  chain forms a closed loop, these 
two quantities are Fourier transform pairs, related via  
$S(f_j)=R(0)+2\sum_{k=1}^{N-1} R(k)\cos(2\pi k f_j)$.
\bibitem{comment2} We did not do the same in the 
calculation of the power spectrum 
because the presence of exponentials in its definition makes 
more time consuming  computational task.
\bibitem{powerlaw} If one assumes that the power laws of 
the PS spans the entire range of frequencies with 
$S(f_j) \sim f_j^{-\beta}$, with $0 < \beta <1$, then, one can 
show analytically that $R(l)$ also presents a power law with 
$R(l) \sim l^{\beta-1}$.
\end{references}
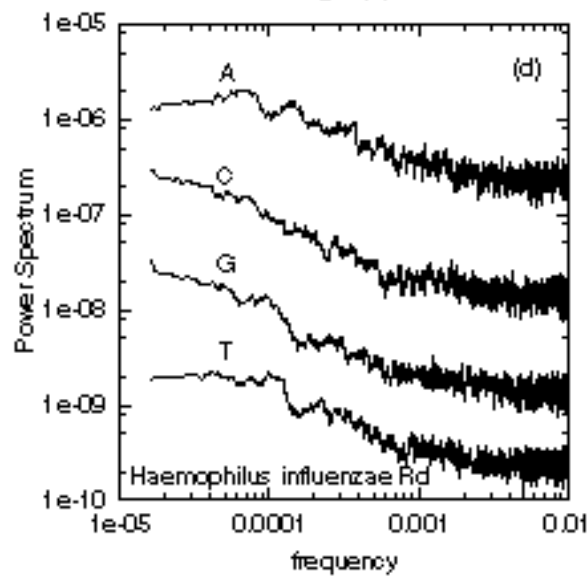
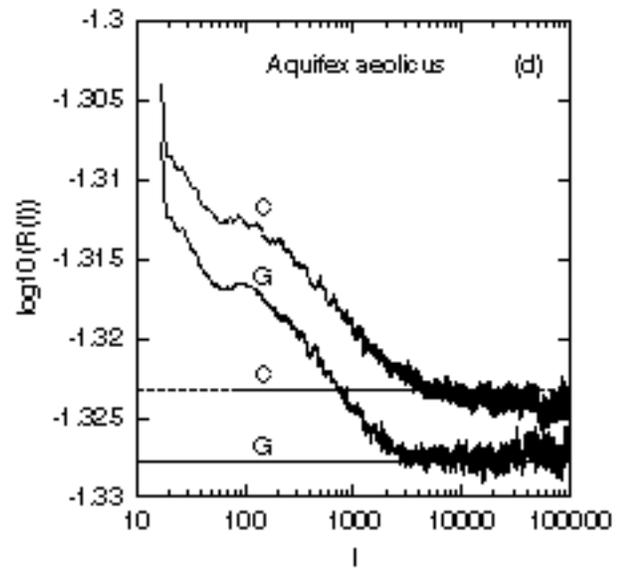
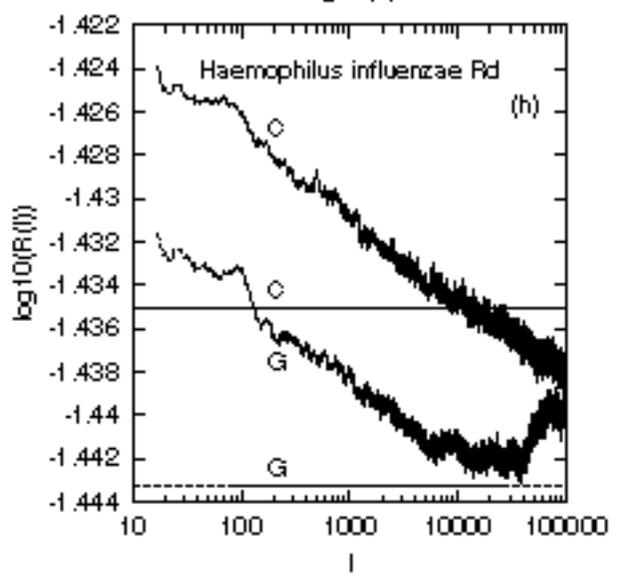
\end{document}